\documentclass[12pt,a4paper]{article}

\usepackage{ifthen} 
\newboolean{pdflatex}
\usepackage[normalem]{ulem}
\setboolean{pdflatex}{true} 

\newboolean{articletitles}
\setboolean{articletitles}{true} 
\newboolean{uprightparticles}
\setboolean{uprightparticles}{false} 

\usepackage{mciteplus}
\usepackage{amsmath}
\usepackage{amssymb}
\usepackage{graphicx}
\usepackage[colorlinks=true, allcolors=blue]{hyperref}
\def\Pieq      {\scalebox{0.8}{$\prod$}}  
\def\pl      {\ensuremath{p_{\ell}^*}\xspace}

\def\B     {\ensuremath{B}\xspace}

\def\Bbar  {\kern 0.18em\overline{\kern -0.18em B}{}\xspace}
\def\Bb    {\ensuremath{\Bbar}\xspace}
\def\Bzb   {\ensuremath{\Bbar^0}\xspace}
\def\Bu    {\ensuremath{B^+}\xspace}

\newcommand {\Bs}{\ensuremath{\Bbar_{s}}\xspace}
\def\Dz    {\ensuremath{D^0}\xspace}
\def\Dstar    {\ensuremath{D^*}\xspace}
\def\Dstarstar    {\ensuremath{D^{**}}\xspace}
\def\Dstarstarz    {\ensuremath{D^{**0}}\xspace}
\def\Ds    {\ensuremath{D^+_s}\xspace}
\def\Dsp   {\ensuremath{D^+_s}\xspace}

\def\Dss   {\ensuremath{D^*_s}\xspace}
\def\Dsstarstar   {\ensuremath{D^{**+}_s}\xspace}

\def\Xc     {\ensuremath{X_c}\xspace}
\def\Xcd     {\ensuremath{X_{cd}}\xspace}
\def\Xcu     {\ensuremath{X_{cu}}\xspace}

\def\mb     {\ensuremath{m_{b}}\xspace}
\def\mc     {\ensuremath{m_{c}}\xspace}

\def\Done  {\ensuremath{D_{1}}\xspace}

\def\Dszero   {\ensuremath{D^{*+}_{s0}}\xspace}
\def\Dsonep   {\ensuremath{D^{'+}_{s1}}\xspace}
\def\Dsone  {\ensuremath{D_{s1}^{+}}\xspace}
\def\Dstwo   {\ensuremath{D^{*+}_{s2}}\xspace}

\def\DszeroNum   {\ensuremath{D^{*}_{s0}(2317)^{+}}\xspace}
\def\DsoneNum  {\ensuremath{D_{s1}(2460)^{+}}\xspace}
\def\DsonepNum  {\ensuremath{D_{s1}(2536)^{+}}\xspace}
\def\DstwoNum   {\ensuremath{D^{*}_{s2}(2573)^{+}}\xspace}

\def\Bsb   {\ensuremath{\Bbar_s^0}\xspace}
 
\def\BzBzb {\ensuremath{B^0 {\kern -0.16em \Bzb}}\xspace}

\def\Bstar       {{\ensuremath{B^{*}}}\xspace}
\def\Bsstar       {{\ensuremath{B_{s}^{*0}}}\xspace}

\newcommand {\mhsq}{\ensuremath{m_{H}^2}}

\newcommand {\Xcs}{\ensuremath{X_{cs}}\xspace}
\newcommand {\Gsls}{\ensuremath{\Gamma_{SL}}}

\newcommand {\vcb}{\ensuremath{|V_{cb}|}\xspace}
\newcommand {\vub}{\ensuremath{|V_{ub}|}\xspace}

\newcommand {\Bxclnu}{\ensuremath{\Bb \to X_c\ellm\bar\nu_\ell}\xspace}

\newcommand{\BsDlnu} {\ensuremath{\Bsb \to D_{s}^{+}\ell^{-}\bar{\nu}_\ell}}
\newcommand{\BsDslnu} {\ensuremath{\Bsb\to D_{s}^{*+}\ell^{-}\bar{\nu}_\ell}}

\newcommand{\BsDszlnu} {\ensuremath{\Bsb\to D_{s0}^{*+}\ell^{-}\bar{\nu}_\ell}}
\newcommand{\BsDsplnu} {\ensuremath{\Bsb\to D_{s1}^{'+}\ell^{-}\bar{\nu}_\ell}}
\newcommand{\BsDspplnu} {\ensuremath{\Bsb\to D_{s1}^{+}\ell^{-}\bar{\nu}_\ell}}
\newcommand{\BsDsdlnu} {\ensuremath{\Bsb\to D_{s2}^{*+}\ell^{-}\bar{\nu}_\ell}}
\newcommand{\BsDsKlnu} {\ensuremath{\Bsb\to D^{(*)} K\ell^{-}\bar{\nu}_\ell}}

\usepackage{xspace} 
\usepackage{upgreek}

\def\lhcb   {\mbox{LHCb}\xspace}

\def\babar  {\mbox{BaBar}\xspace}
\def\belle  {\mbox{Belle}\xspace}

\def\delphi {\mbox{DELPHI}\xspace}

\def\lhc    {\mbox{LHC}\xspace}

\def\MagUp {\mbox{\em Mag\kern -0.05em Up}\xspace}

\ifthenelse{\boolean{uprightparticles}}%
{
 
 \def\Pgamma      {\ensuremath{\upgamma}\xspace}

 \def\Pmu         {\ensuremath{\upmu}\xspace}                 
 \def\Pnu         {\ensuremath{\upnu}\xspace}                 
                  
 \def\Ppi         {\ensuremath{\uppi}\xspace}

 \def\PDelta      {\ensuremath{\Delta}\xspace}                 
 \def\PXi         {\ensuremath{\Xi}\xspace}                 
 \def\PLambda     {\ensuremath{\Lambda}\xspace}                 
 \def\PSigma      {\ensuremath{\Sigma}\xspace}                 
 \def\POmega      {\ensuremath{\Omega}\xspace}                 
 \def\PUpsilon    {\ensuremath{\Upsilon}\xspace}

 \def\PB      {\ensuremath{\mathrm{B}}\xspace}                 
                  
 \def\PD      {\ensuremath{\mathrm{D}}\xspace}

 \def\PK      {\ensuremath{\mathrm{K}}\xspace}

 \def\PZ      {\ensuremath{\mathrm{Z}}\xspace}                 
                  
 \def\Pb      {\ensuremath{\mathrm{b}}\xspace}                 
 \def\Pc      {\ensuremath{\mathrm{c}}\xspace}

 \def\Pi      {\ensuremath{\mathrm{i}}\xspace}

 \def\Ps      {\ensuremath{\mathrm{s}}\xspace}

 \def\thebaroffset{0.0em}
}
{
 
 \def\Pgamma      {\ensuremath{\gamma}\xspace}

 \def\Pmu         {\ensuremath{\mu}\xspace}                 
 \def\Pnu         {\ensuremath{\nu}\xspace}                 
                  
 \def\Ppi         {\ensuremath{\pi}\xspace}

 \mathchardef\PDelta="7101
 \mathchardef\PXi="7104
 \mathchardef\PLambda="7103
 \mathchardef\PSigma="7106
 \mathchardef\POmega="710A
 \mathchardef\PUpsilon="7107
                  
 \def\PB      {\ensuremath{B}\xspace}                 
                  
 \def\PD      {\ensuremath{D}\xspace}

 \def\PK      {\ensuremath{K}\xspace}

 \def\PZ      {\ensuremath{Z}\xspace}                 
                  
 \def\Pb      {\ensuremath{b}\xspace}                 
 \def\Pc      {\ensuremath{c}\xspace}

 \def\Pi      {\ensuremath{i}\xspace}

 \def\Ps      {\ensuremath{s}\xspace}

 \def\thebaroffset{0.18em}
}
\newcommand{\offsetoverline}[2][\thebaroffset]{\kern #1\overline{\kern -#1 #2}}%

\makeatletter
\ifcase \@ptsize \relax
  \newcommand{\miniscule}{\@setfontsize\miniscule{4}{5}}
\or
  \newcommand{\miniscule}{\@setfontsize\miniscule{5}{6}}
\or
  \newcommand{\miniscule}{\@setfontsize\miniscule{5}{6}}
\fi
\makeatother

\DeclareRobustCommand{\optbar}[1]{\shortstack{{\miniscule (\rule[.5ex]{1.25em}{.18mm})}
  \\ [-.7ex] $#1$}}

\def\mun        {{\ensuremath{\Pmu^-}}\xspace} 

\def\ellm       {{\ensuremath{\ell^-}}\xspace}
\def\ellp       {{\ensuremath{\ell^+}}\xspace}

\def\neu        {{\ensuremath{\Pnu}}\xspace}

\def\g      {{\ensuremath{\Pgamma}}\xspace}

\def\Z      {{\ensuremath{\PZ}}\xspace}

\def\squark    {{\ensuremath{\Ps}}\xspace}

\def\cquark    {{\ensuremath{\Pc}}\xspace}

\def\bquark    {{\ensuremath{\Pb}}\xspace}

\def\pion   {{\ensuremath{\Ppi}}\xspace}
\def\piz    {{\ensuremath{\pion^0}}\xspace}
\def\pip    {{\ensuremath{\pion^+}}\xspace}
\def\pim    {{\ensuremath{\pion^-}}\xspace}

\def\kaon    {{\ensuremath{\PK}}\xspace}

\def\KorKbar {\kern \thebaroffset\optbar{\kern -\thebaroffset \PK}{}\xspace}
\def\Kz      {{\ensuremath{\kaon^0}}\xspace}

\def\Kp      {{\ensuremath{\kaon^+}}\xspace}

\def\KS      {{\ensuremath{\kaon^0_{\mathrm{S}}}}\xspace}

\def\D       {{\ensuremath{\PD}}\xspace}

\def\DorDbar {\kern \thebaroffset\optbar{\kern -\thebaroffset \PD}\xspace}
\def\Dz      {{\ensuremath{\D^0}}\xspace}

\def\Dp      {{\ensuremath{\D^+}}\xspace}
\def\Dm      {{\ensuremath{\D^-}}\xspace}

\def\DpDm    {\ensuremath{\Dp {\kern -0.16em \Dm}}\xspace}
\def\Dstar   {{\ensuremath{\D^*}}\xspace}

\def\Dstarz  {{\ensuremath{\D^{*0}}}\xspace}

\def\Dstarp  {{\ensuremath{\D^{*+}}}\xspace}
\def\Dstarm  {{\ensuremath{\D^{*-}}}\xspace}

\def\Ds      {{\ensuremath{\D^+_\squark}}\xspace}
\def\Dsp     {{\ensuremath{\D^+_\squark}}\xspace}

\def\Dss     {{\ensuremath{\D^{*+}_\squark}}\xspace}
\def\Dssp    {{\ensuremath{\D^{*+}_\squark}}\xspace}

\def\B       {{\ensuremath{\PB}}\xspace}
\def\Bbar    {{\ensuremath{\offsetoverline{\PB}}}\xspace}
\def\Bb      {{\ensuremath{\Bbar}}\xspace}
\def\BorBbar {\kern \thebaroffset\optbar{\kern -\thebaroffset \PB}\xspace}

\def\Bzb     {{\ensuremath{\Bbar{}^0}}\xspace}
\def\Bd      {{\ensuremath{\B^0}}\xspace}
\def\Bdb     {{\ensuremath{\Bbar{}^0}}\xspace}
\def\BdorBdbar {\kern \thebaroffset\optbar{\kern -\thebaroffset \Bd}\xspace}
\def\Bu      {{\ensuremath{\B^+}}\xspace}

\def\Bs      {{\ensuremath{\B^0_\squark}}\xspace}
\def\Bsb     {{\ensuremath{\Bbar{}^0_\squark}}\xspace}
\def\BsorBsbar {\kern \thebaroffset\optbar{\kern -\thebaroffset \Bs}\xspace}

\def\Y#1S{\ensuremath{\PUpsilon{(#1S)}}\xspace}

\def\FiveS {{\Y5S}}

\def\Lz          {{\ensuremath{\PLambda}}\xspace}

\def\LorLbar     {\kern \thebaroffset\optbar{\kern -\thebaroffset \PLambda}\xspace}

\def\Lb           {{\ensuremath{\Lz^0_\bquark}}\xspace}

\newcommand{\decay}[2]{\ensuremath{#1\!\to #2}\xspace} 

\def\to                 {\ensuremath{\rightarrow}\xspace}

\def\qsq       {{\ensuremath{q^2}}\xspace}

\def\Vcb  {{\ensuremath{V_{\cquark\bquark}}}\xspace}

\def\AT#1     {\ensuremath{A_{\mathrm{T}}^{#1}}\xspace}           

\def\C#1      {\ensuremath{\mathcal{C}_{#1}}\xspace}                       
\def\Cp#1     {\ensuremath{\mathcal{C}_{#1}^{'}}\xspace}                    
\def\Ceff#1   {\ensuremath{\mathcal{C}_{#1}^{\mathrm{(eff)}}}\xspace}        
\def\Cpeff#1  {\ensuremath{\mathcal{C}_{#1}^{'\mathrm{(eff)}}}\xspace}       
\def\Ope#1    {\ensuremath{\mathcal{O}_{#1}}\xspace}                       
\def\Opep#1   {\ensuremath{\mathcal{O}_{#1}^{'}}\xspace}                    

\newcommand{\bra}[1]{\ensuremath{\langle #1|}}             
\newcommand{\ket}[1]{\ensuremath{|#1\rangle}}              

\newcommand{\aunit}[1]{\ensuremath{\text{\,#1}}}       

\newcommand{\tev}{\aunit{Te\kern -0.1em V}\xspace}
\newcommand{\gev}{\aunit{Ge\kern -0.1em V}\xspace}
\newcommand{\mev}{\aunit{Me\kern -0.1em V}\xspace}
\newcommand{\kev}{\aunit{ke\kern -0.1em V}\xspace}
\newcommand{\ev}{\aunit{e\kern -0.1em V}\xspace}
 
\newcommand{\mevc}{\ensuremath{\aunit{Me\kern -0.1em V\!/}c}\xspace}
\newcommand{\gevc}{\ensuremath{\aunit{Ge\kern -0.1em V\!/}c}\xspace}
\newcommand{\mevcc}{\ensuremath{\aunit{Me\kern -0.1em V\!/}c^2}\xspace}
\newcommand{\gevcc}{\ensuremath{\aunit{Ge\kern -0.1em V\!/}c^2}\xspace}

\def\gsim{{~\raise.15em\hbox{$>$}\kern-.85em
          \lower.35em\hbox{$\sim$}~}\xspace}
\def\lsim{{~\raise.15em\hbox{$<$}\kern-.85em
          \lower.35em\hbox{$\sim$}~}\xspace}

\def\tell1  {TELL1\xspace}
\def\ukl1   {UKL1\xspace}

\newcommand{\eg}{\mbox{\itshape e.g.}\xspace}

\begin{document}
\begin{titlepage}

\begin{flushright}
{\small
Nikhef 2023-022\\
\today \\
}
\end{flushright}

\vskip1cm
\begin{center}
{\Large \bf\boldmath Inclusive semileptonic $\Bs$ meson decays at the LHC via a sum-of-exclusive modes technique: possibilities and prospects}
\end{center}

\vspace{0.5cm}
\begin{center}
{\sc M. De Cian$^{a}$, N. Feliks$^{b,\dag}$, M. Rotondo$^{c}$ and K. Keri Vos$^{d,e}$} \\[3mm]

{\it $^a$Physikalisches Institut,\\ Ruprecht-Karls-Universit\"at Heidelberg,\\ D-69120 Heidelberg, Germany
}\\[0.3cm]

{\it $^b$Institute of Physics,\\ Ecole Polytechnique F\'ed\'erale de Lausanne (EPFL),\\ CH-1015 Lausanne, Switzerland \\ $^{\dag}$ Author was at institute at time work was performed.
}\\[0.3cm]

{\it $^c$INFN Laboratori Nazionali di Frascati,\\
Frascati, Italy
}\\[0.3cm]

{\it $^d$Gravitational 
Waves and Fundamental Physics (GWFP),\\ 
Maastricht University, Duboisdomein 30,\\ 
NL-6229 GT Maastricht, the
Netherlands}\\[0.3cm]

{\it $^e$Nikhef, Science Park 105,\\ 
NL-1098 XG Amsterdam, the Netherlands}
\end{center}

\vspace{0.6cm}
\begin{abstract}

\vskip0.2cm\noindent
We propose an approach for measuring the moments of the hadronic invariant mass distribution in semileptonic \Bs meson decays using a sum-of-exclusive technique. Using the present and foreseen knowledge about exclusive semileptonic \Bs decays, we estimate the uncertainties on moments of the kinematic distribution. Semileptonic \Bs decays can be described, as their \Bd counterpart, using the Heavy Quark Expansion (HQE), with the only difference between the \Bs and \Bd mesons being the $SU(3)_F$ breaking effects that change the numerical values of the non-perturbative HQE parameters. We extract the HQE parameters for the \Bs decays from our estimates of the moments, showing the potential of the proposed method. We identify a set of required measurements for a future precision measurement.

\end{abstract}
\end{titlepage}


\section{Introduction}
Thanks to impressive theoretical and experimental advances, the semileptonic $b\to c \ell \bar\nu_\ell$ decay allows for a precise extraction of the CKM-matrix element \vcb. 
This fundamental parameter of the Standard Model of Particle Physics (SM) can be extracted from two type of decays, depending on the hadronization of the final state: from {\it exclusive} final states such as $B\to D$ and $B\to \Dstar$, and from \textit{inclusive} final states, denoted by $B\to \Xc$, where \Xc sums over all possible final states with a \cquark quark. The determinations of \vcb ~from inclusive and exclusive decays have shown a consistent discrepancy since about 15 years (see \eg~\cite{Gambino:2020jvv} for a recent review). 
Inclusive decays are described by an operator product expansion (OPE) using the heavy-quark expansion (HQE). This framework allows the calculation of the semileptonic decay width, which depends on a set of non-perturbative parameters that can be extracted from the decay distributions and specifically from kinematic moments of the spectrum. High-precision measurements of these moments have been performed at the B factories, but experiments at hadron colliders can give additional information, as shown by the CDF measurement of the hadronic mass moments~\cite{CDF:2005xlh}. Advances in the HQE framework allow for a precision of $1-2\%$ in extractions of $|\Vcb|$ \cite{Bordone:2021oof, Fael:2018vsp, Finauri:2023kte}. 

In this paper, we explore the feasibility of a measurement of the hadronic mass distribution of inclusive $\Bsb\to X_{cs} \ellm \bar{\nu}_\ell$ decays \footnote{Throughout the paper the inclusion of charge-conjugate decay modes is implied. The sign conventions apply to $b\to c$ transitions, but can be extended to ${\bar b}\to {\bar c}$ transitions with proper sign change.}, based on a sum-of-exclusive technique. This measurement could \eg be performed at the \lhcb experiment, or any other hadron-collider experiment with similar capabilities. 
Using \Bs mesons instead of \Bd or \Bu mesons has the advantage of having well-separated resonances in the charm meson spectrum. This is advantageous from an experimental point of view to determine their yields, as it avoids performing an amplitude analysis to determine interference effects, as would be the case for the spectrum of \Xcd or \Xcu mesons. 
At high-energy hadron colliders such as the \lhc, \Bs\ mesons are abundantly produced, with a production fraction of about 25\% compared to \Bd mesons at the \lhcb experiment, opening the possibility of precise studies of their semileptonic decays. 

From a theoretical point of view \Bs mesons offer several new possibilities. A measurement of moments of the spectrum could constrain the HQE parameters of the semileptonic \Bs\ decay for the first time, allowing to quantify the amount of $SU(3)_F$ violation and to improve existing estimates of the total semileptonic width of the \Bs\ meson~\cite{Bigi:2011gf,Bordone:2022qez}. Better knowledge of these parameters, and in particular of the Darwin term $\rho^3_D(\Bs)$, could improve the prediction of the total \Bs\ meson decays width, currently based on $SU(3)_F$ estimates of this parameter~\cite{King:2021jsq, Lenz:2021bkv}. A more precise measurement of the semileptonic width of the \Bs\ meson would also improve the determination of the production fraction of \Bs, $f_s$, at LHCb ~\cite{LHCb:2019fns}, whose uncertainty enters in many analyses involving \Bs mesons. The experimental analysis would also lead to a better understanding of the \Dsstarstar states and thereby to better estimates of the universality ratio in these excited states, $R(\Dsstarstar)$ 
~\cite{Bernlochner:2016bci, Bernlochner:2017jxt,Bernlochner:2021vlv} . And finally, the procedure described in this paper, may open the road for a possible \vcb determination from \Bs decays by combining moments of the spectrum with the total semileptonic decay rate, which could play a role in resolving the discrepancy between inclusive and exclusive measurements. 

The paper is outlined as follows. We first discuss the Heavy Quark Expansion and present estimates for the HQE parameters in \Bs decays. In Sec.~\ref{sec:previous_studies}, we briefly mention previous studies at high-energy colliders. In Sec.~\ref{sec:expknow}, we discuss the available information on \Bs decays. We then discuss the challenges and present a proof-of-concept. In Sec.~\ref{sec:momana}, we present an estimate for the moments and their uncertainty based on current data and a projection for the future. Here we also extract the HQE parameters from the experimental data. Finally, we conclude in Sec.~\ref{sec:conc}.

\section{Heavy Quark Expansion}
The description of inclusive decays relies on the Heavy Quark Expansion, which allows to express the triple differential rate as an expansion in inverse powers of the heavy-quark mass \mb with perturbatively calculable coefficients. Here we consider the inclusive \Bs decay:
\begin{equation}
  \Bsb(p_B)\to \Xcs(p_X) \ellm(p_\ell) \bar{\nu}_\ell(p_\nu) \ ,
\end{equation}
where \Xcs contains all final states with an $s$ and $c$ quark and where we define its hadronic mass $m_H=M(\Xcs)$. In addition, $E_\ell$ is the lepton energy and $(p_\ell + p_\nu)^2 = \qsq$ the dilepton invariant mass squared. 

For $b\to c\ell\bar{\nu}_\ell$ decays, the OPE is usually set up by treating the \cquark quark as a heavy degree of freedom assuming $\mb \sim \mc \gg \Lambda_{\rm QCD}$. In this case, we obtain the standard $1/\mb$ expansion for the total rate expressed in local hadronic matrix elements
\begin{align}\label{eq:hqe}
    d\Gsls & = d\Gamma_0 + \left(\frac{\Lambda_{\rm QCD}}{m_b}\right)^2 d\Gamma_2 
    + \left(\frac{\Lambda_{\rm QCD}}{m_b}\right)^3 d\Gamma_3
    + \left(\frac{\Lambda_{\rm QCD}}{m_b}\right)^4 d\Gamma_4 \\
    & +  \left[a_0\left(\frac{\Lambda_{\rm QCD}}{m_b}\right)^5 + a_1 \left(\frac{\Lambda_{\rm QCD}}{m_b}\right)^3\left(\frac{\Lambda_{\rm QCD}}{m_c}\right)^2\right] d\Gamma_5 + \ldots \ .
\end{align}
Here
\begin{equation}
    d\Gamma_i = \sum_n C_i^{(n)}\bra{\Bsb}  \mathcal{O}_i^{(n)} \ket{\Bsb} \ ,
\end{equation}
where $C_i^{(n)}$ are short-distance coefficients and $\mathcal{O}$ are operators with mass dimension $i+3$, with $n$ running over all operators at the given order in the expansion. At leading order, $d\Gamma_0$ presents the partonic rate, $d\Gamma_1$ vanishes due to the equation of motions and starting at $d\Gamma_2$ non-perturbative forward matrix elements, the HQE parameters, enter. 

Specifically, we consider terms up to $1/\mb^3$ defined as (see \eg ~\cite{Mannel:2010wj})\footnote{The differences between this basis and those with full covariant derivatives are given in the Appendix of \cite{Fael:2018vsp} (see also \cite{Mannel:2023yqf}).}: 
\begin{eqnarray}
2 m_{\Bs} \, (\mu_\pi^2)^\perp & \equiv & 
- \langle \Bs |\bar b_v (iD_\rho) (iD_\sigma) b_v| \Bs \rangle \Pieq^{\rho\sigma} \ ,
\notag
\\
2 m_{\Bs} \, (\mu_G^2)^\perp & \equiv & \frac{1}{2}
\langle \Bs |\bar b_v \left[i D_\rho, i D_\lambda\right] (- i \sigma_{\alpha \beta}) b_v| \Bs \rangle  \Pi^{\alpha \rho}  \Pieq^{\beta\lambda},
\notag
\\
2 m_{\Bs} \, (\rho_D^3)^\perp & \equiv & \frac{1}{2}
\langle \Bs |\bar b_v \left[i D_\rho, \left[iD_\sigma, iD_\lambda\right]\right] b_v| \Bs \rangle \Pi^{\rho\lambda}  v^\sigma,
\notag
\\
2 m_{\Bs} \, (\rho_{LS}^3)^\perp & \equiv & \frac{1}{2}
\langle \Bs |\bar b_v\left\{i D_\rho, \left[iD_\sigma, iD_\lambda\right]\right\}(-i\sigma_{\alpha\beta}) b_v| \Bs \rangle \Pieq^{\alpha\rho}\Pieq^{\beta\lambda}  v^\sigma\, ,
\label{eqn:HQEm2}
\end{eqnarray}
where $v^\mu = p_B^\mu/m_{\Bs}$ is the velocity of the \Bs meson and
\begin{equation}
    \Pieq_{\mu\nu} = g_{\mu\nu}-v_{\mu}v_\nu \ . 
\end{equation}
Higher order terms have also been identified, but we will not consider those here \cite{Mannel:2010wj,Gambino:2016jkc, Mannel:2023yqf}. The setup of the OPE is such that the matrix elements depend on the mass of the initial-state \Bs meson. Therefore there is a difference between the HQE parameters defined for \Bs and \Bd decays, which arises due to $SU(3)_F$ corrections. For simplicity, we do not label the matrix elements and also drop the ``perp'' superscript in the following.

The HQE predicts the triple-differential rate in terms of the HQE parameters. However, the predictions only hold for sufficiently inclusive quantities and differential rates cannot be compared point by point with the data. Therefore only kinematic moments of the differential spectrum, like the leptonic invariant mass \qsq, the charged-lepton energy $E_\ell$ or the hadronic invariant mass (squared) $M_X^2$ can be precisely calculated. In this paper, we focus on moments of the hadronic invariant mass defined as
\begin{equation}\label{eq:mxdef}
    m_H^2 \equiv m_{\Bs}^2 + q^2 - 2 m_{\Bs}\ v\cdot q \ ,
\end{equation}
where $v^\mu = p_{\Bs}^\mu/m_{\Bs}$ is the velocity of the \Bs meson. Moments of the differential spectrum are then defined as
\begin{equation}\label{eq:hadmom}
    M_n = \langle {(\mhsq)}^n \rangle = \int{ (\mhsq)^{n} \frac{1}{\Gsls}\frac{d\Gsls}{d\mhsq}} d\mhsq.\\
\end{equation}
Due to the experimental setup, these moments usually depend on kinematical cuts, either in the lepton energy or in the \qsq momentum. We discuss this in more detail below. Finally, we will consider centralized moments defined through
\begin{equation}\label{eq:momcen}
    M'_n = \langle ({\mhsq}-\langle {\mhsq} \rangle)^n \rangle = \int{ (\mhsq-M_1)^{n} \frac{1}{\Gsls}\frac{d\Gsls}{d\mhsq}} d\mhsq.\\
\end{equation}

The $m_H^2$ moments are thus combinations of \qsq and $v\cdot q$ moments. In these the mass of the spectator quark does not enter explicitly, therefore the perturbative corrections to these moments are the same as in the \Bd case. However, since the mass of the heavy meson enters explicitly in the definition of the hadronic invariant mass, the expressions will be different from their \Bd counterparts. In addition, as discussed before, the HQE elements also differ.

The HQE for inclusive $\bar{B}\to \Xc$ decays has seen quite some progress in the recent years, where now also corrections up to $\alpha^3_s$ for the total rate \cite{Fael:2020tow} are available. For the $M_X$ moments, $\alpha_s^2$ corrections are available for fixed \mb and \mc values and for fixed lepton energy cuts \cite{Pak:2008cp,Pak:2008qt,Gambino:2011cq}. In Sec.~\ref{sec:momex}, we give equations for the first three centralized moments. To avoid renormalon issues with the short-distance pole mass, we employ the kinetic mass for the $b$ quark \cite{Bigi:1996si} and the $\overline{\rm MS}$ mass for the \cquark quark. For our numerical analysis, we take \cite{Fael:2020njb,Fael:2020iea}
\begin{equation}\label{eq:mbmc}
    \mb^{\rm kin}(1 \;{\rm GeV}) =  4.565 \pm 0.020\; {\rm GeV} \ , \quad\quad    \overline{m}_c(2\; \rm{GeV}) = 1.093 \pm 0.008\; \rm{GeV}  \ .
\end{equation}

With these inputs, we can write a simple numerical formula for the branching ratio\footnote{Which holds for both \Bd and \Bs decays, and only differs through the values of the HQE parameters.}
\begin{align}
    \mathcal{B}(\Bsb\to \Xcs\ellm\bar{\nu}_\ell)&= 66.85 |V_{cb}|^2\left[\left(1-\frac{\mu_\pi^2}{2\mb^2}\right) -0.40 \alpha_s^2 -0.17 \alpha_s^3 -1.91 \frac{\mu_G^2}{\mb^2} \right. \nonumber \\
    {}& + 
 \left.  \alpha_s \left(-0.39+0.36 \frac{\mu_\pi^2}{\mb^2}\right)
-16.68 \frac{\rho_D^3}{\mb^3} +1.91 \frac{\rho_{LS}^3}{\mb^3}\right]
\end{align}

\subsection{\boldmath Difference between \Bd and \Bs from $SU(3)_F$}
\label{sec:diffbs}
Measurements of the $M_X$ or other moments would allow the extraction of the HQE elements in \Bs decays, and allow an $SU(3)_F$ analysis based on the experimental data. We stress that it would be very interesting to compare those with direct determinations using lattice QCD. Recently, a lot of progress has been made in this direction both for \Bd and \Bs mesons (see \eg~\cite{Gambino:2020crt, Gambino:2017vkx,Gambino:2022dvu, Barone:2023tbl,Gambino:2023xoe}). Estimates of the size of the $SU(3)_F$ breaking effects in the HQE are discussed in detail in \cite{Bigi:2011gf}, which was recently updated and extended in \cite{Bordone:2022qez}. We also refer to \cite{Kobach:2019kfb} for a similar discussion on the $SU(3)_F$ effects.

The chromomagnetic operator $\mu_G^2$ can be extracted from the hyperfine splitting of the $B_{(s)}$ meson masses. We have 
\begin{equation}
(m^2_{\Bsstar}-m^2_{\Bs}) = \frac{4}{3}C_{\rm mag}\mu_G^2(\Bs)+ \mathcal{O}(1/\mb)   \ ,
\end{equation}
where $C_{\rm mag}$ accounts for higher-order corrections. Taking, $C_{\rm mag}=1$ and using the masses from \cite{Workman:2022ynf} gives
\begin{equation}\label{eq:muges}
    \mu_G^2(\Bs) = (0.40 \pm 0.08)\; {\rm GeV}^2 \ ,
\end{equation}
where we have increased the uncertainty to account for $20\%$ effects from neglected $1/\mb$ suppressed terms. A similar calculation for the \Bd meson gives $\mu_G^2(\Bs) = (0.36 \pm 0.07)$. In \cite{Bordone:2022qez}, also lattice determinations of the mass difference \cite{Gambino:2017vkx} were considered. Averaging those with the hyperfine splitting measurements gives for the ratio 
\cite{Bordone:2022qez}
\begin{equation}
        \frac{\mu_G^2(\Bs)}{\mu_G^2(\Bd)} = 1.14 \pm 0.10 \ ,
\end{equation}
which indicates only a $14\%$ $SU(3)_F$ breaking in this parameter. Combining this with the recent value of \cite{Bordone:2021oof} for $\mu_G^2(\Bd) = 0.306$ GeV$^2$ obtained from a global fit to the $B\to \Xc$ semileptonic data\footnote{We note that the extracted value from experimental data is about one standard deviation lower than the estimate from the hyperfine splitting.} , gives 
\begin{equation}\label{eq:mugcons}
\mu_G^2(\Bs) = (0.35 \pm 0.07)\; {\rm GeV}^2 \ ,
\end{equation}
which is in agreement with our estimate in \eqref{eq:muges}. We take this as our ``SM'' estimate for this parameter. 

Similarly, the analysis in \cite{Bordone:2022qez} finds
\begin{equation}
    \mu_\pi^2(\Bs) - \mu_\pi^2(\Bd) = (0.1 \pm 0.1)\; {\rm GeV}^2 \ ,
\end{equation}
which using $\mu_\pi^2(\Bd) = 0.477  $ GeV$^2$, results in 
\begin{equation}
    \mu_\pi^2(\Bs) = (0.58 \pm 0.10)\; {\rm GeV}^2.
\end{equation}
 Finally, for $\rho_D^3$, an estimate of the $SU(3)_F$ contributions can be made using a vacuum saturation approximation linking it to the ratio of decay constants $f_{\Bs}/f_{\Bd}$ or using heavy-quark sum rules. As in \cite{Bordone:2022qez}, we take the average of both
 \begin{equation}\label{eq:rholscons}
     \frac{\rho_D^3(\Bs)}{\rho_D^3(\Bd)} \simeq 1.39 \pm 0.15 \ ,
 \end{equation}
 which using $\rho_D^3(\Bd) = (0.185\pm 0.03)$ GeV$^3$ \cite{Bordone:2021oof} gives
 \begin{equation}
     \rho_D^3(\Bs) \simeq (0.26 \pm 0.03) \; {\rm GeV}^3 \ .
 \end{equation}

Finally, we note that $\rho_{LS}^3$ has only a very small influence on the total rate\footnote{This is because $\rho_{LS}^3$ is not a reparametrization-invariant quantity and can be reabsorbed into $
\mu_G^2$ (see \eg~\cite{Mannel:2018mqv}). } and therefore the $SU(3)_F$ corrections to this parameter were not discussed in \cite{Bordone:2022qez}. As an estimate, we take the obtained value in \Bd decays with a slightly increased uncertainty to account for $30\%$ $SU(3)_F$ breaking added in quadrature to the extracted value
\begin{equation}
    \rho_{LS}^3(\Bs) \simeq -(0.13 \pm 0.10) \; {\rm GeV}^3
\end{equation}

In the next parts, we use these above values to get estimates for the ``SM'' predictions for the $M_X$ moments.

\section{Previous experimental studies of inclusive decays at high-energy colliders}
\label{sec:previous_studies}
Experiments at hadron colliders, or in conditions where the \B meson does not originate from a resonance with a known mass, lack the possibility to fully determine the event kinematics for (inclusive) semileptonic decays. Despite this limitation, several determinations of hadronic mass moments have been performed at the Tevatron and at LEP.

In the measurement by the CDF collaboration\cite{CDF:2005xlh} the final hadronic state \Xc in the \Bxclnu\ inclusive decays is split into three contributions, corresponding to the \Dz, \Dstarz and \Dstarstarz mesons. The \Dstarstarz meson here includes any neutral excited charm state decaying into $\Dstar\pi$ or $\D\pi$ mesons, in a resonant or non-resonant manner. The measurement of the hadronic moments requires the determination of the relative ratio of the \Dstarstar component to the \Dstar and \D components, and the determination of the shape of the moments of the $\Dstar\pi$ or $\D\pi$. 

A similar approach has been used by the DELPHI collaboration ~\cite{DELPHI:2005mot}. 
In their measurement, the boost of the \B mesons produced at energies around the \Z peak ensures sensitivity to the full lepton spectrum without a truncation of the lepton energy, thus reducing the model dependence. The CDF measurement cannot access the full lepton spectrum, but corrections based on Monte Carlo studies were introduced to ensure a threshold at $\pl > 0.7\gevcc$, where \pl is the momentum of the lepton in the centre-of-mass frame of the \B meson.

Since the CDF and DELPHI measurements, further experimental knowledge on the semileptonic decays of \B mesons into excited \Dstarstar mesons has been accumulated. In particular a decay of the excited \Done meson into the $\D\pip\pim$ final state has been observed by the Belle experiment~\cite{Belle:2004bvv}, while the \babar experiment  observed the semileptonic decay \decay{\Bb}{\Dstar\pip\pim \ellm\bar\nu_\ell}~\cite{BaBar:2015zkb}, which has been confirmed with higher precision by Belle~\cite{Belle:2022yzd}. These contributions, which account for about $5-10\%$ of the total inclusive semileptonic rate, would need to be included in the evaluation of future hadronic moment measurements when using the same approach as CDF and DELPHI. Further known contributions to consider would be the decays of $B^{-}$ mesons into the $D_{s}^{(*)-} \Kp\ellm\bar\nu_\ell$ final state, measured by both \babar \cite{BaBar:2010ner} and Belle \cite{Belle:2012ccr}. In addition, theoretically allowed but yet unmeasured decays (\eg $B^{-}\to \Lambda_c^- p \ellm\bar\nu_\ell$ or $B^-\to D^{0(*)}\eta \ellm\bar\nu_\ell$) are expected to contribute to the inclusive spectra, although their impact on the final result is likely negligible, taking the limited event yield and systematic effects of the CDF and DELPHI measurements into account.

\section{Current experimental knowledge of semileptonic \boldmath{\Bs} meson decays}\label{sec:expknow}

Analogous to semileptonic decays of \Bu and \Bd mesons,  the charm meson states \Xcs\ in the decay \decay{\Bsb}{\Xcs \ellm \bar\nu_\ell} can be split in \Ds, \Dssp, \Dsstarstar, where \Dsstarstar stands for any excited state with a mass larger than the \Dssp meson, and non-resonant contributions.  

The semileptonic \Bs decays are not known very well, with only a few measurements being performed so far. The branching fractions of the \decay{\Bsb}{\Dsp\mun\bar\neu_{\mu}} and \decay{\Bsb}{\Dssp\mun\bar\neu_{\mu}} decays, which are expected to be the leading contribution to \Xcs, have only been recently measured by the \lhcb collaboration \cite{LHCb:2020cyw}. These decays are known with a total uncertainty of about $10\%$.

One of the challenges in the study of \Bs semileptonic decays into  the \Dsstarstar excited states is the poor knowledge of the branching fractions of these states. 
The  dominant contribution to the \Dsstarstar mesons are the four well-established $P$-wave states \Dszero, \Dsonep, \Dsone and \Dstwo\footnote{In the literature, these four states are also denoted as \DszeroNum, \DsoneNum, \DsonepNum and \DstwoNum~
\cite{Workman:2022ynf}.}. The two lightest $L=1$ \Ds excited states, \Dszero\ and \Dsonep, have a small decay width. They have a mass below the $\Dz\Kp$ and $\Dstarz\Kp$ mass threshold, respectively, so they only decay via the strong interaction to the $\Ds\pi$ or $\Dssp\pi$ final states, or via electromagnetic processes.  Since their discovery,  several studies have been done on these mesons, because they are good candidates to be tetraquarks or bound $D^{(*)}K$ states.  At present, the only observed decay of the \Dszero meson, \decay{\Dszero}{\Dss\piz}, has been measured by the BESIII collaboration, with an uncertainty on the branching fraction of 20\% \cite{BESIII:2017vdm}. The other allowed decay modes, \decay{\Dszero}{\Dss\g} and \decay{\Dszero}{\Ds\g\g}, have not been observed yet.  Several final states of the \Dsonep\ meson decay have been observed, many involving neutral particles. The most precise measurements were performed with the \Dssp\piz and \Dsp\g final state, both with a relative uncertainty of about 22\% \cite{Workman:2022ynf}.
 
The masses of the \Dsone and \Dstwo states are larger than the threshold for $\Dz\Kp$ and $\Dstar\Kp$ production, so they preferentially decay to the $\Dstar\Kp$ or $\Dz\Kp$ final states, mainly via $D$-wave processes. Because these states have masses close to the threshold, their natural width is very narrow. 
No absolute branching fraction results exist for the \Dsone and \Dstwo states: for the \Dsone meson, \Dstarp\Kz and \Dstarz\Kp final states appear most dominant, while for the \Dstwo meson, the \Dz\Kp and \Dp\Kz final states have been observed \cite{Workman:2022ynf}.
 
The mass, width and principal decay modes of the 4 $P$-wave states are summarized in Table~\ref{tab:Ds}. 

\begin{table}
\centering
\caption{\label{tab:Ds} Masses and widths of the excited states and their decay modes~\cite{Workman:2022ynf}. Only measured branching fractions and upper limits are reported, ``seen'' is used when a decay is established, but no branching fraction has been reported. Note that for the \Dsone meson, the decay to \Dstarz\Kp is defined as 100\%, and all other branching fraction are measured relative to it.}
\vspace{0.2cm}
\begin{tabular}{ l r | l r | l r | l r }
\hline\hline
     \multicolumn{2}{c|}{$\Dszero$} & 
     \multicolumn{2}{|c|}{$\Dsonep$} & 
     \multicolumn{2}{|c|}{$\Dsone$} & 
     \multicolumn{2}{|c}{$\Dstwo$} \\\hline

    \multicolumn{2}{c|}{$2317.8\pm0.5\mev$} & 
    \multicolumn{2}{|c|}{$2459.5\pm 0.6\mev$} & 
    \multicolumn{2}{|c|}{$2535.11\pm 0.06\mev$} & 
    \multicolumn{2}{|c}{$2569.1\pm 0.8\mev$} \\ 
     
    \multicolumn{2}{c|}{$<3.8\mev$} & 
    \multicolumn{2}{|c|}{$<3.5\mev$} & 
    \multicolumn{2}{|c|}{$0.92\pm0.05\mev$} & 
    \multicolumn{2}{|c}{$16.9\pm0.7\mev$}\\\hline
    
    $\Ds\pi^0$  & $100^{+0}_{-20}$\% &
    $\Dss \pi^0$& $48\pm 11$\% &    
    $D^{*+}\KS$ & $85\pm 12$\% & 
    $D^0 K^+$   & seen  \\   
    
    $\Ds\gamma$   & $<5$\% &
    $\Ds\gamma$   & $18\pm 4$\% &    
    $D^{*0} K^+$  & 100\% & 
    $\Dp\KS$  &  seen \\ 
    
    $\Dss\gamma$    & $<6$\% &
    $\Ds\pip\pim$  & $4.3\pm 1.3$\% &    
    $D^+\pim \Kp$  & $2.8\pm 0.5$\% &
    $\Dstarp\KS$ & seen \\ 
    
    $\Ds\gamma\gamma$ & $<18$\%   &
    $\Dss\gamma$   & $<8$\% &    
    $\Ds\pip\pim$ & seen &
    &  \\
    
    &  &
    $\Dszero\gamma$ & $3.7^{+5.0}_{-2.4}$\% &    
    $\Dp\Kz$    &  $<34$\% &
    &  \\

    &  &
    &  &    
    $\Dz\Kp$    & $ <12$\% &
    &  \\
    \hline
    
\end{tabular}
\end{table}

The small natural widths of these four states, and consequently the lack of interference between any two states, make the \Xcs\ spectrum  qualitatively very different to the \Xc\ spectrum from \B\ decays. As mentioned, this presents a significant advantage from an experimental point of view, as the inclusive decay width can be treated as the sum of exclusive resonant components.

Besides the states described so far, higher mass \Ds states exist as well~\cite{Workman:2022ynf}, they are expected to predominantly decay to \Dz\Kp,  \Dp\Kz or $DK\pi$ combinations.  These have been observed in the study of the $DK$ mass spectra in \decay{\Bs}{DK\pi} and \decay{\B}{DK\pi} hadronic decays, and in the $DK\pi$  mass spectrum of \decay{\Bd}{D DK\pi} decays \cite{LHCb:2020gnv}.  

In the following we briefly summarize the present knowledge, or expectations, on the \Bs semileptonic decays into \Ds excited $L=1$ states:

\begin{enumerate}

\item \decay{\Bsb}{\Dszero\ellm\bar\neu_{\ell}} and \decay{\Bsb}{\Dsonep\ellm\bar\neu_{\ell}}: these decays have not been observed yet. Since the discovery of the \Dszero and \Dsonep mesons, many calculations of the \decay{\Bsb}{\Dszero} and \decay{\Bsb}{\Dsonep} form factors have been reported. The predicted branching fractions are in the range of $0.1\%$ to $0.4\%$, \cite{Huang:2004et,Aliev:2006gk, Aliev:2006qy, Faustov:2012mt,Navarra:2015iea,Kang:2018jzg,Gubernari:2023rfu}. In the following study we assume a branching fraction of $0.3\%$ for both \decay{\Bsb}{\Dszero\ellm\bar\neu_{\ell}} and \decay{\Bsb}{\Dsonep\ellm\bar\neu_{\ell}} decays, with a relative uncertainty of $100\%$.

\item \BsDspplnu :  this decay mode has been observed by the D0 collaboration using the \decay{\Dsonep}{\Dp \KS} decay \cite{D0:2007ukf}, and by \lhcb using the \decay{\Dsonep}{\Dz \Kp} decay mode \cite{LHCb:2011rmd}. Considering the \DsonepNum decay modes shown in Table \ref{tab:Ds}, we estimate ${\cal {B}}(\BsDspplnu) = (0.98\pm 0.20) \%$. 

\item \BsDsdlnu : this decay has been observed by \lhcb using the \decay{\Dstwo}{\Dz\Kp} decay mode \cite{LHCb:2011rmd}. Taking into account the \Dstwo decay modes shown in Table \ref{tab:Ds}, we estimate ${\cal {B}}(\BsDsdlnu) = (0.58\pm 0.20) \%$.

\end{enumerate}

We summarize the current knowledge of the semileptonic branching fractions of the \Bs meson (measured or estimated) in the first column (Conf. A) in Table~\ref{tab:BsSemileptonic}. Conf. B is a future scenario that we discuss in the next section.

\begin{table}[t!]
\centering
\caption{\label{tab:BsSemileptonic}
The branching fractions of the contributions to the $\Bsb\to X_{cs}\ellm\bar{\nu}_\ell$ decay. Configuration (Conf.) A reflects the current knowledge on the branching ractions, Conf. B an improved precision that could be achieved in the future, as explained in the text.}
\vspace{0.2cm}
\begin{tabular}{ l r | r}
\hline\hline
$B_s^0$ Decay      & $\cal{B}$[\%] (Conf. A)   &  $\cal{B}$[\%] (Conf. B) \\
\hline
$\Bsb\to X_{cs}\ell\bar\nu_\ell$                        & 10.05$\pm$0.31 & 10.05$\pm$0.31 \\ 
\hline
\BsDlnu  ~\cite{LHCb:2020cyw}           &  2.44$\pm$0.23   & 2.44$\pm$ 0.10 \\
\BsDslnu ~\cite{LHCb:2020cyw}           &  5.3$\pm$0.5     & 5.30 $\pm$0.22 \\
\BsDszlnu  ~(see text)                                    &  0.3$\pm$0.3     & 0.30$\pm$0.03 \\ 
\BsDsplnu  ~(see text)                                    &  0.3$\pm$0.3     & 0.30$\pm$0.03 \\
\BsDspplnu                      &  0.98$\pm $0.20  & 0.98$\pm$0.05 \\
\BsDsdlnu                       &  0.58$\pm $0.20  & 0.58$\pm$0.04 \\  \hline\hline    
\BsDsKlnu ~(see text)                      &  0.15$\pm $0.15  & 0.150$\pm$0.015\\
\hline

\end{tabular}

\end{table}

The semileptonic \Bs decays into higher mass excited states have not been observed yet, but their predicted branching fractions are below $0.1\%$, as shown in Refs.~\cite{Gan:2014jxa, Alomayrah:2022lne}.  Above the $DK$ mass threshold, one would expect the non-resonant contribution of the \BsDsKlnu~  decays. This has been observed to be non-negligible, as shown by \lhcb in \cite{LHCb:2019fns}. Therefore, for the following studies, the branching ratio of the \BsDsKlnu~ decay is estimated using the difference between the theory prediction of the full semileptonic decay width of the \Bs meson and the sum of all aforementioned decay widths. For the branching ratio, we use the predicted ratio of the semileptonic decay widths of the \Bs and \Bd mesons in \cite{Bordone:2022qez} given by 
\begin{equation}
    \Gamma_{SL}(\Bs)/\Gamma_{SL}(\Bd)=1-(0.018\pm 0.008).
\end{equation}
Using the ratio of the \Bs and \Bd lifetimes and the \Bd branching ratio from \cite{Workman:2022ynf} and assuming that the \vub-suppressed $\Bsb\to X_u\ellm\bar\nu_\ell$ rate is the same as for the \Bd meson, we find 
\begin{equation}\label{eq:brbs}
     \mathcal{B}(\Bsb \to \Xc\ellm \bar{\nu}_\ell)=(10.05\pm 0.31)\%,
\end{equation}
where the uncertainty is dominated by the uncertainty on the inclusive $\Bdb\to X\ellm\bar\nu_\ell$ branching fraction. Finally, we then get the branching fractions for the \BsDsKlnu ~modes as 
\begin{equation}
    \mathcal{B}(\BsDsKlnu ) = \mathcal{B}(\Bsb\to X_{cs}\ellm \bar{\nu}_\ell) - \Sigma_{\rm res}  \mathcal{B}({
    \rm res}).
\end{equation}
This leads to $\mathcal{B}(\BsDsKlnu ) = 0.15\pm 0.15$ as given in Table~\ref{tab:BsSemileptonic}, where we assign a $100\%$ relative uncertainty to this ratio.

\begin{figure}[!t]
\centering
    \includegraphics[width=0.9\textwidth]{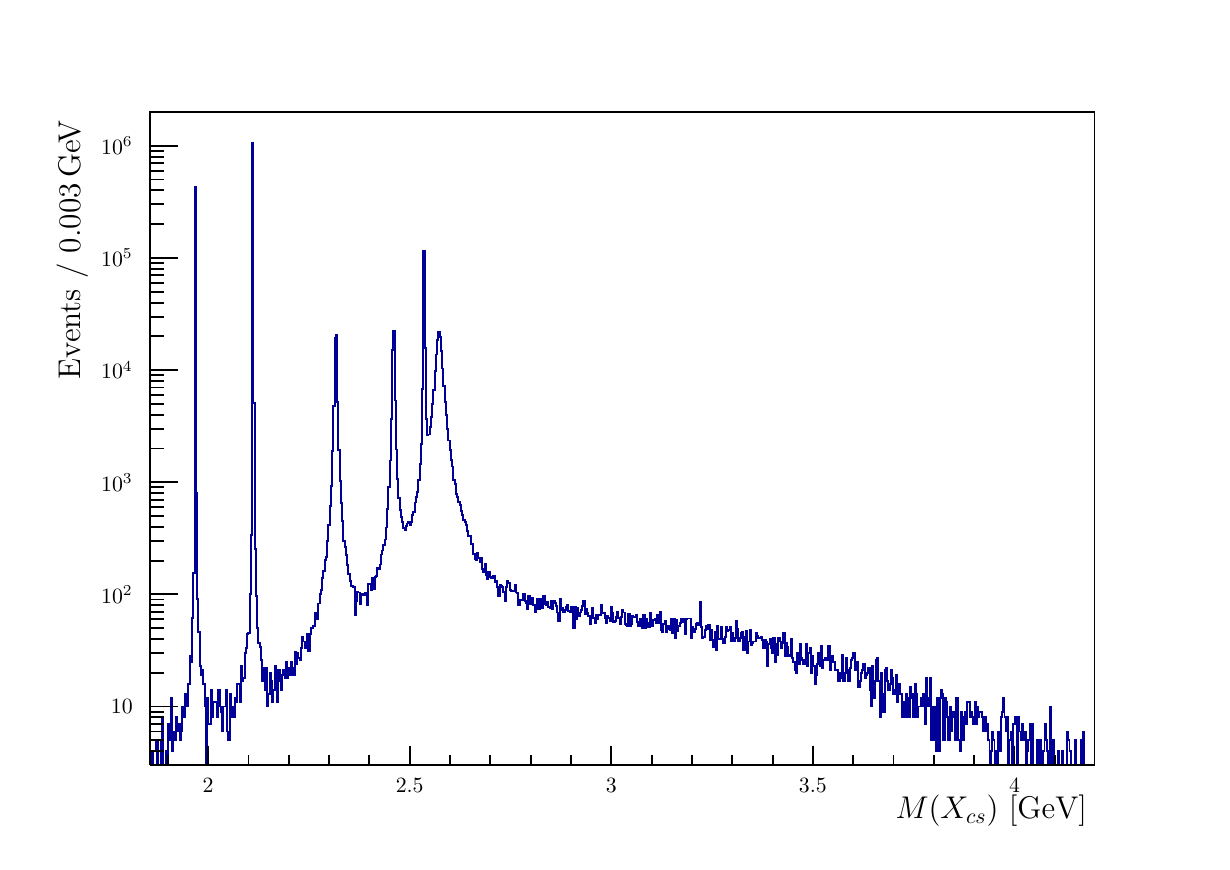}
    \caption{Distributions of $M(X_{cs})$ of a pseudo-experiment. The peaks correspond to the resonant contributions.}
    \label{fig:xc_mass}
\end{figure}

\section{Proof-of-concept}\label{sec:proof}
In order to prove the validity of the method, we extract hadronic mass moments from a simulated \Xcs spectrum. To obtain the spectrum, we use the values of the masses and the widths of the charm resonances in Table~\ref{tab:Ds}, and those of the \Dsp and \Dssp mesons from \cite{Workman:2022ynf} combined with the branching fractions given in Table~\ref{tab:BsSemileptonic}. 
Above the $DK$ mass threshold, we consider the presence of the non-resonant  \BsDsKlnu~ decays. Other possible non-resonant contributions are ignored in the present study. A list of some of those are:
\begin{enumerate}
\item \decay{\Bsb}{\Ds\piz\ellm\bar\nu_\ell} and \decay{\Bsb}{\Ds\pip\pim\ellm\bar\nu_\ell}: these decays are expected to be OZI suppressed.
\item \decay{\Bsb}{\Ds\eta\ellm\bar\nu_\ell} and \decay{\Bsb}{\Ds\phi\ellm\bar\nu_\ell}: these processes are the analogous to the \decay{\Bdb}{\Ds^{(*)} K\ellm\bar\nu_\ell} decay observed at $B$ Factories. For \B meson decays this contribution is about $0.6\%$ of the semileptonic width, so these decays are expected to also be very small for \Bs semileptonic decays;
\item \decay{\Bsb}{D^{(*)}K(n\pi)\eta\ellm\bar\nu_\ell}: decays with emissions of one or more pions. These contributions should be suppressed by the limited phase space available to the decays and are ignored. 
\end{enumerate}

Because of the narrowness of both $L=0$ and $L=1$ $D_{s}^{(*,**)+}$ states, we can write the semileptonic differential \mhsq\ spectrum, normalized to the total semileptonic rate, as
\begin{equation}\label{eq:diffdistr}
    \frac{1}{\Gsls}\frac{d\Gsls}{d\mhsq}=\sum_{L=0}\frac{\Gamma_i}{\Gsls}\cdot \delta(\mhsq-m_i^2) + \sum_{L=1}\frac{\Gamma_i}{\Gsls}\cdot \delta(\mhsq-m_i^2)+ \frac{\Gamma_{DK}}{\Gsls}\cdot f^{DK}(\mhsq),\\
\end{equation}
where \Gsls\ is the inclusive \Bs\ semileptonic width in \eqref{eq:hqe}, $\Gamma_i$ are the decay widths of the \decay{\Bsb}{ \D_{s}^{(*,**)+}\ellm\bar\nu_\ell} decays, $m_i^2$ are mass-squared values of the various $\Ds^{(*,**)}$ states, $f^{DK}(\mhsq)$ is the normalized hadronic mass-squared distribution for the \decay{\Bsb}{DK\ellm\bar\nu_\ell} decays, beyond the $P$-wave resonances. 

The hadronic mass moments of this spectrum in \eqref{eq:hadmom}, can then be obtained via the equation
\begin{equation}
    M_{n}=\sum_{L=0}\frac{\Gamma_i}{\Gsls}\cdot {(m_i^2)^n} + \sum_{L=1}\frac{\Gamma_i}{\Gsls}\cdot {(m_i^2)^n} + \frac{\Gamma_{DK}}{\Gsls}\cdot M_{n}^{DK},\\
\end{equation}
where $M_{n}^{DK}$ are the moments of the $f^{DK}(\mhsq)$ distribution.

In order to obtain the spectrum, 
we generate large set of pseudo-experiments, assuming that the hadronic state in the inclusive \decay{\Bsb}{\Xcs \ellm \bar\nu_\ell} is constituted by the resonant states with $L=0$ and $L=1$ shown in Tab.~\ref{tab:BsSemileptonic}, and the non-resonant \BsDsKlnu\ decays. In each pseudo-experiment the branching ratios of each decay mode are generated independently, with Gaussian distributions. All resonances were simulated using a Breit-Wigner lineshape.  

In order to describe the lineshape $f^{DK}(\mhsq)$ of the non-resonant  \decay{\Bsb}{\Dz\Kp\ellm\bar\nu_\ell} contribution, a modification of the so-called Goity-Roberts model is used~\cite{PhysRevD.51.3459}. The original Goity-Roberts model as used in EvtGen~\cite{Lange:2001uf} describes the emission of a soft pion in semileptonic \Bd or \Bu decays. For this work the formalism was slightly modified to describe the emission of a kaon from semileptonic \Bs decays by using the mass splitting between the \Bsstar and \Bs mesons (48.5\mev) instead of the one between the \Bstar and \Bd mesons (45.2\mev), and by using the kaon decay constant (155.7\mev) instead of the pion decay constant (93\mev). An experimentally driven model for the decay \decay{\Bsb}{\Dz\Kp\mun\bar\nu_\mu} has been used in \cite{LHCb:2019fns}, however no functional form has been published and therefore it is not considered for this work.

In Fig.~\ref{fig:xc_mass} the distribution of the \Xcs\ mass spectrum for a single pseudo-experiment is shown as an example. No experimental resolutions or efficiencies are taken into account for this distribution. 

We simulate the mass spectrum of \Xcs\ with two configurations: Configuration (Conf.) A  uses the current uncertainties on these modes according to Tab.~\ref{tab:BsSemileptonic}.
Conf. B presents a future scenario where these branching ratios are known more precisely. For this configuration, we assume that the uncertainty on the branching fractions of \BsDlnu\ and \BsDslnu\ will be reduced to $5\%$ using the full dataset collected by \lhcb during Run 1 and 2 of the \lhc. The \BsDszlnu\ and \BsDsplnu\ decays have not been observed yet, but we assume they could be measured with uncertainties close to $10\%$. These uncertainties can be achieved by improving the knowledge of the branching fractions of the \Dsone meson which, at present, are not known well.
The uncertainties on the branching fractions of the \BsDspplnu\ and \BsDsdlnu\ decays are assumed to become of the order of $5\%$ and $7\%$, respectively. Also in this case it would be paramount to improve the knowledge on the \Dsonep and \Dszero decay modes. 
For the non-resonant component, we again assume that the $DK$ saturates the total rate, and that it will be known with a $10\%$ uncertainty.

\section{Moment Analysis}
\label{sec:momana}

\vspace{0.2cm}
\begin{table}[!t]
    \centering
          \caption{Extracted values of the moments for configuration A and B. The correlations between the moments are also reported. In the last column the moments are extracted assuming only the contributions of the resonant $L=0$ and $L=1$ states.}
\vspace{0.2cm}
    \begin{tabular}{c|c c c | c c c| c}
    \hline
    Moments 			  & 	Conf. A 	  & $M'_2$ &   $M'_3$    & Conf. B & $M'_2$ & $M'_3$  & $L=0$ and $L=1$\\
    \hline
    $M_1$ [$\rm{GeV}^2$]  &   4.82$\pm$ 0.08  & 0.74 & 0.55     &4.78$\pm$0.02 & 0.72 & 0.45 & 4.79$\pm$0.02\\ 
    $M'_2$ [$\rm{GeV}^4$] &   1.36 $\pm$ 0.29 &      & 0.96     &1.22$\pm$0.05 & & 0.90 & 0.82$\pm$0.09\\
    $M'_3$ [$\rm{GeV}^6$] &   4.7 $\pm$ 1.8   &      &          &3.86$\pm$0.28 & & & 1.07$\pm$0.11\\    
    \hline
    \hline
    \end{tabular}
            \label{tab:mom}
\end{table}

\begin{figure}[!t]
\centering
    \includegraphics[width=1.0\textwidth]{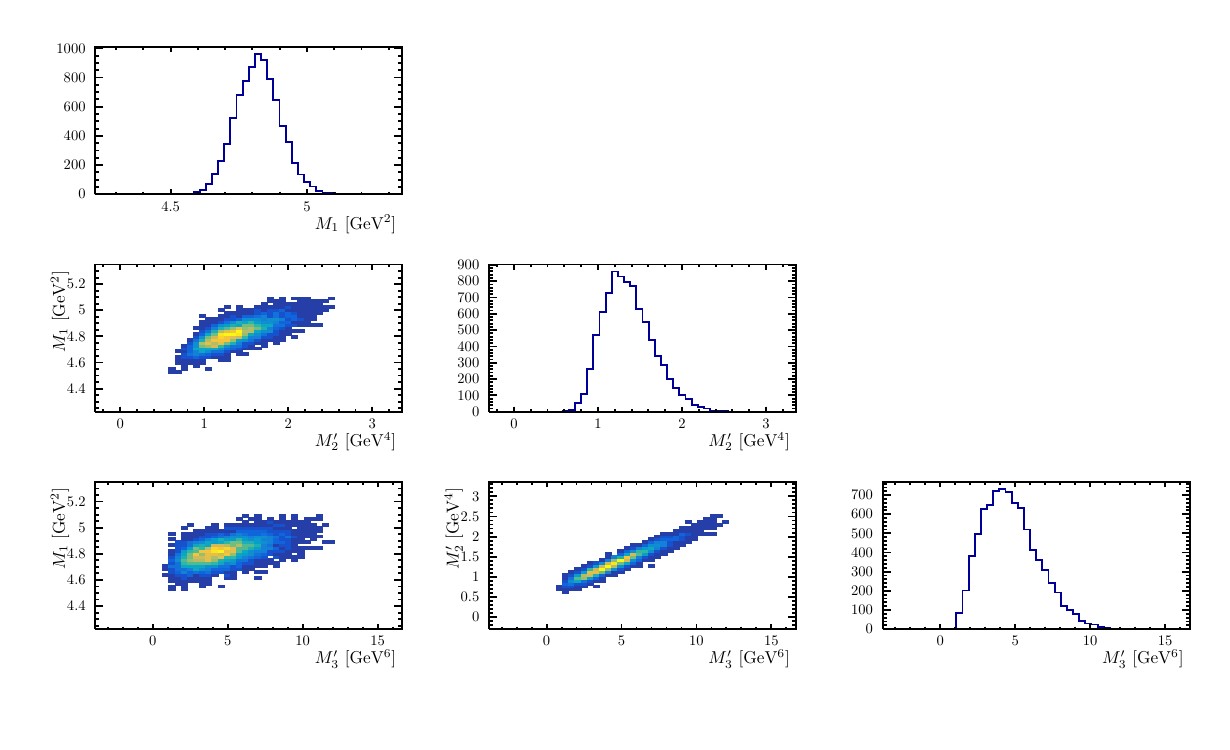}
    \caption{Distributions of the three $M_x$ moments (diagonal) and their correlations obtained in Conf. A.}
    \label{fig:mom_corr}
\end{figure}

The distribution of the moments determined from the pseudo-experiments are shown in Fig.~\ref{fig:mom_corr}. It can be noted that the two higher-order central moments $M'_2$ and $M'_3$, have a correlation above $90\%$. The resulting moments, and their correlations, are reported in Tab.~\ref{tab:mom} for both configurations considered. 
For comparison we report also the moments assuming only the resonant states. While the moment $M_1$ is only marginally affected by the non-resonant component, the higher-order moments, $M'_2$ and $M'_3$, depend crucially on this component.

\subsection{Expressions for the moments and theoretical inputs}\label{sec:momex}
In this section, we discuss what the extracted experimental moments imply for the HQE parameters. For this study, we fix both the mass of the \bquark quark and the \cquark quark as in \eqref{eq:mbmc}. 
With these inputs, we can find easy expressions for the centralized moments in \eqref{eq:momcen} in terms of the HQE parameters. Without a kinematic cut, we find 
\begin{align}
    M_1 &= 4.85 + 0.30 \alpha_s + 0.46 \frac{\mu_G^2}{\rm GeV^2} - 0.68 \frac{\mu_\pi^2}{\rm GeV^2} +  0.99 \frac{\rho_D^3}{\rm GeV^3} - 0.12 \frac{\rho_{LS}^3}{\rm GeV^3} \ , \nonumber \\
    M'_2 &= 0.28 + 1.47 \alpha_s -0.30 \frac{\mu_G^2}{\rm GeV^2} +4.77 \frac{\mu_\pi^2}{\rm GeV^2} -6.0 \frac{\rho_D^3}{\rm GeV^3}+0.28 \frac{\rho_{LS}^3}{\rm GeV^3} \ , \nonumber \\
       M'_3 &= -0.058 +3.3 \alpha_s + 0.04 \frac{\mu_G^2}{\rm GeV^2} +3.6 \frac{\mu_\pi^2}{\rm GeV^2} + 23.96 \frac{\rho_D^3}{\rm GeV^3}+0.96 \frac{\rho_{LS}^3}{\rm GeV^3} \ ,
\end{align}
in units of GeV$^2$, GeV$^4$ and GeV$^6$, respectively. These expressions differ from the ones in \cite{Gambino:2004qm} due to $m_{\Bs}$ entering the definition of the $M_X$ moments in \eqref{eq:mxdef}. These expressions were checked against the Koyla-package that is under development \cite{kolya}. In this first study, we do not include the $\alpha_s^2$ corrections. These corrections are known analytically without a kinematic cut \cite{Pak:2008cp,Pak:2008qt} (see also \cite{Fael:2022frj} for a first study up to $\alpha_s^3$), and mix with $\alpha_s$ corrections from the scheme transformation of the quarks and kinetic parameters. We note that not all lepton energies may be accessible and that in fact a cut on the lepton energy would be required. In the future, 
it is worth to investigate if also moments as a function of the lepton energy cut can be obtained, as done for the \Bd analysis. At the moment, we use the above expressions without any kinematical constraints. 

To get an idea of the numerical values of the moments, we can make ``SM'' estimates for these moments, using the inputs defined in Sec.~\ref{sec:diffbs} for the \Bs decays based on the $SU(3)_F$ assumptions. We find
\begin{align}
    M_1^{SU(3)_F} \simeq (4.95 {}&\pm 0.08)\; {\rm GeV}^2 \ , \quad (M'_2)^{SU(3)_F} \simeq (1.67 \pm 0.52) \;{\rm GeV}^4 \ , \nonumber \\
    {}& (M'_3)^{SU(3)_F}\simeq (8.80 \pm 0.84)\; {\rm GeV}^6 \ ,  
\end{align}
where the uncertainty comes from the HQE inputs and a variation of $\alpha_s = 0.22 \pm 0.07$ to account for missing higher perturbative terms. We note here that \cite{Fael:2022frj} studied the $\alpha_s^3$ corrections, which were found to be large for the third moment. This point therefore requires further theoretical studies. We do not include an uncertainly for missing higher-order HQE terms and note that half of the uncertainty of the third moment comes from the uncertainty of $\rho_D^3$, highlighting the sensitivity to this parameter.

For comparison, we also quote the value of the moments without kinematic cut for the \Bd decays
\begin{align}
    M_1(\Bd) = (4.45 {}& \pm 0.08)\; {\rm GeV}^2 \ , \quad M'_2(\Bd) = (1.56\pm 0.50)\; {\rm GeV}^4 \ , \nonumber \\
    {}& M'_3(\Bd) = (6.20 \pm 0.78)\; {\rm GeV}^6 \ .  
\end{align}
As expected, we see that especially the third centralized moment is very sensitive to the $SU(3)_F$ breaking effects and high-order HQE elements. Comparing with the estimated moments in Table~\ref{tab:mom}, we note a large difference in the third moment. This could be due to over-estimating the $\rho_D^3$ input value or due to missed higher resonances (or non-resonant contributions) in the experimental determination.

Finally, we compare these predictions with the \Bd measurements of the \delphi collaboration, which also do not have any kinematic cut. They quote \cite{DELPHI:2005mot}
\begin{align}
    M_1(\Bd) = (4.54 \pm {}&0.10)\; {\rm GeV}^2 \ ,   \quad M'_2(\Bd) = (1.56\pm 0.24)\; {\rm GeV}^4 \ , \nonumber \\
    {}& M'_3(\Bd)=(4.05 \pm 0.81)\; {\rm GeV}^6  \ ,
\end{align}
where we note (as is known) that the third moment is quite off from the theoretical predictions that employs HQE elements extracted from this data. We quote these values to get a sense of the uncertainty available in \Bd decays. From \babar, only $M_X$ moments are available with a lepton energy cut. Taking the values at $E_\ell = 1.1$ GeV, we find\footnote{These are obtained from the normal moments including the correlated uncertainties. We also note that the measurements at even lower lepton energy cuts have very large uncertainties. } \cite{BaBar:2009zpz}
\begin{align}
    M_1(\Bd) = (4.36 \pm {}&0.07)\; {\rm GeV}^2 \ ,   \quad M'_2(\Bd) = (1.06\pm 0.22)\; {\rm GeV}^4 \ , \nonumber \\
    {}& M'_3(\Bd)=(3.0 \pm 1.2)\; {\rm GeV}^6  \ ,
\end{align}

We note that the lepton energy cut significantly reduces 
the third moment.

 \subsection{Extracting the \boldmath{\Bs} HQE elements from data}
We can now fit our extracted moments in Table~\ref{tab:mom} to obtain the \Bs elements directly from data. Before we proceed, we note that the HQE parameters obtained from this fit should be taken with caution. In order to get an idea of the sensitivity for the HQE parameters, we take here the extracted moments as presented in Table~\ref{tab:mom}. We do not account for any additional uncertainty due to the assumptions on the modelling of higher resonances. Therefore, the extracted HQE parameters should be seen as a proof of principle and not as actual determinations of these parameters.

To set up our fit, we follow a similar strategy as for fits in the \Bd system (see \cite{Finauri:2023kte, Gambino:2004qm, Bernlochner:2022ucr}). First, we use the constraint from the hyperfine splitting in \eqref{eq:mugcons} to constrain $\mu_G^2$. In addition, we also constrain $\rho_{LS}^3$ using \eqref{eq:rholscons}, which is the same constraint used in the \Bd analyses. We leave $\rho_D^3$ and $\mu_\pi^2$ free in the fit. 

The experimental correlations between the $M_X$ moments are given in Table~\ref{tab:mom}. We note quite a large correlation between the moments, which is not surprising given our sum-over-exclusives setup. For the theoretical uncertainties, we have to account for missing higher orders both on the perturbative and non-pertubative side. To account for the first, we vary $\alpha_s = 0.22\pm 0.07$. As in the \Bd analysis, we account for missing higher-order terms in the HQE expansion by varying both $\rho_D^3$ by $30\%$ and $\mu_G^2$ by $20\%$. For this study, we take these uncertainties between different moments completely uncorrelated. This estimate for the theoretical uncertainties may be conservative, given the small effect of higher-order terms (see \eg~\cite{Gambino:2016jkc}). 

Clearly, a disadvantage at this point is that we have only $M_X$ moments available, while analogous \Bd fits are performed with lepton energy moments and/or $q^2$ moments. In addition, measurements at several kinematic cuts are available for the \Bd case. Even though the theoretical predictions are expected to be highly correlated among different cuts (and so are the measurements), having this larger data set would reduce the uncertainty on the extracted parameters.

With the caveats mentioned above, we obtain, 
\begin{equation}
     \mu_\pi^2 = (0.46 \pm 0.12)\; {\rm GeV}^2 \ , \quad \rho_D^3 = (0.16 \pm 0.06)\; {\rm GeV}^3 \ ,
\end{equation}
which are lower than the expected values based on $SU(3)_F$. In addition, we note that these parameters are smaller than those extracted from \Bd decays. We find
\begin{equation}
    \frac{\mu_\pi^2(\Bs)}{\mu_\pi^2(\Bd)} \sim 0.96\ , \quad\quad\quad \frac{\rho_D^3(\Bs)}{\rho_D^3(\Bd)} \sim 0.86\ ,
\end{equation}
where we used the values for the \Bd HQE parameters from \cite{Bordone:2021oof}. Here we do not quote an uncertainty, because of all the caveats mentioned before. In fact, this extraction is quite surprising compared to the $SU(3)_F$ estimates and thus highlights the importance of further investigating moment extractions of the \Bs decays.
For the parameters we constrain, we find
\begin{equation}
\mu_G^2 = (0.33\pm 0.07)\; {\rm GeV}^2 \ , \quad \rho_{LS}^3 = -(0.12 \pm 0.10)\; {\rm GeV}^3 \ ,
\end{equation}
which are very close to the constraints we put in, especially for $\rho_{LS}^3$ showing the reduced sensitivity to this parameter.

\begin{figure}[!t]
\centering
    \includegraphics[width=1.0\textwidth]{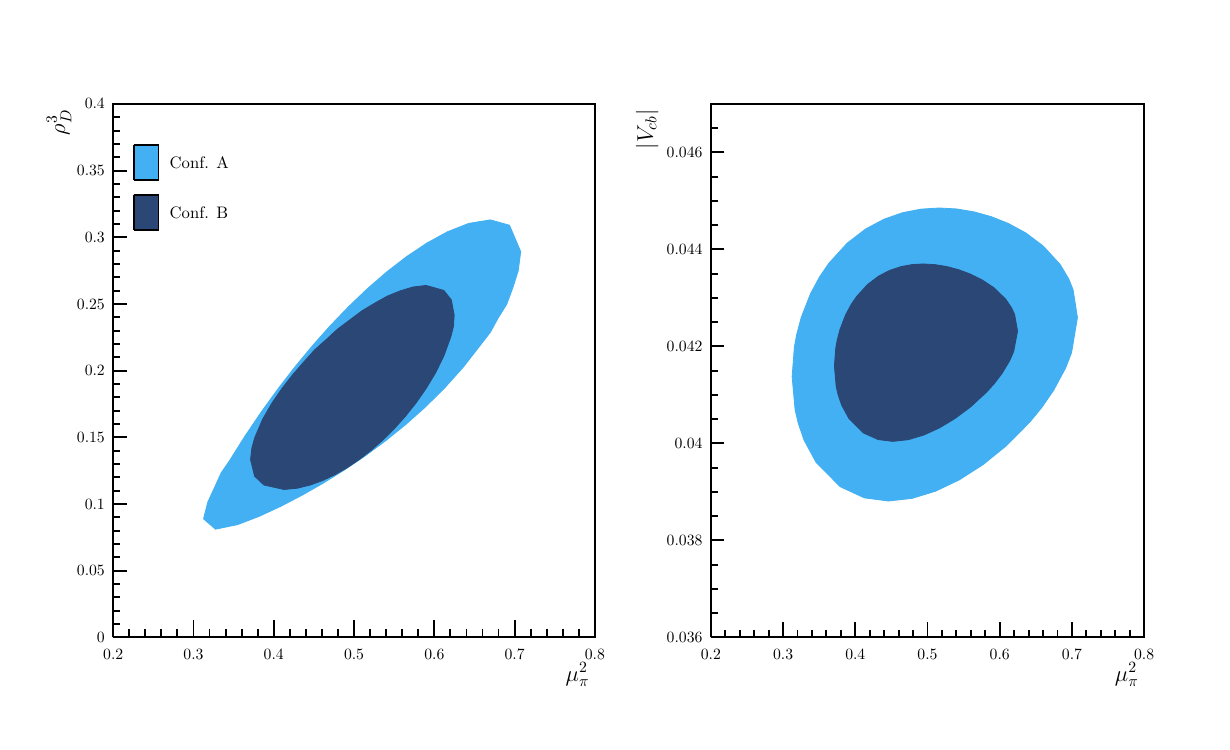}
    \caption{Correlation between the extracted HQE parameters (in units of GeV to the appropriate power) and \vcb for both Conf. A and B.}
    \label{fig:cc}
\end{figure}

In Fig.~\ref{fig:cc}, we give the correlations between $\rho_D^3$ and $\mu_\pi^2$. We note a strong correlation between these two parameters, which was also noted explicitly in \cite{Finauri:2023kte}.

At this point, we again comment on the extracted values of $\rho_D^3$ and $\mu_\pi^2$. These values should clearly not be taken at the same level as those for its \Bd counterpart, they simply present an estimate of what the current sum-over-exclusive \Bs data combined with a model for the higher-mass resonances tells us. The small values for $\rho_D^3$ and $\mu_\pi^2$ extracted from the moments are mainly due to the low value of $M_3'$ compared to the theoretical prediction. As mentioned before, this could be due to underestimating higher mass resonances in the experimental data for which we do not assign an additional uncertainty.

This study presents a first effort to explore the possibilities for future analyses of the \Bs decays modes using a sum-over-exclusive approach, with the extracted values indicating where to improve over the status quo. 
To emphasize this, we use the moments of the future configuration B in Table~\ref{tab:mom} to perform the same fit. Using the same procedure to estimate the theoretical uncertainties, we obtain 
\begin{equation}
     \mu_\pi^2 = (0.47 \pm 0.07)\; {\rm GeV}^2 \ , \quad \rho_D^3 = (0.16 \pm 0.04)\; {\rm GeV}^3 \ ,
\end{equation} 
where the uncertainties are dominated by the theory. 

In this context it could be interesting to study if a soft cut on the hadronic mass $M(\Xcs)< M_X^{\rm cut} \approx 2.8-3.0~\gev$ could be applied. Recently, a study for inclusive $B\to X_s\ellp^+\ellm^-$ decays in~\cite{Huber:2023qse}, shows only a limited breakdown of the OPE with such a minimal mass cut. 

Finally, we may also extract a value for \vcb using experimental input for the branching ratio. While clearly not meant as a precise determination, it serves as a proof-of-concept and as a reference what future studies of semileptonic \Bs decays could achieve.
A measurement of the absolute branching fraction is unlikely to be performed at a hadron-collider experiment, therefore a \vcb determination also requires external input on the branching ratio, for example measured by the \belle II collaboration with dedicated runs at the \FiveS. 
Pioneering measurements of the \Bs inclusive semileptonic branching fraction were obtained by \babar \cite{BaBar:2011sxq} and \belle \cite{Belle:2012mwf}. The averaged branching fraction, dominated by the \belle measurements, is \cite{Workman:2022ynf}

\begin{equation}
{\cal B}(\Bsb \to X\ellm \bar{\nu}_\ell)=(9.6 \pm 0.8)\%, \\
\end{equation}

\noindent which is consistent with the indirect estimation in \eqref{eq:brbs}. 
Using this branching fraction, ignoring the suppressed $b\to u$ contribution, we find 
\begin{equation}
     \vcb = (41.8 \pm 2.0)\cdot 10^{-3}   \ .
\end{equation}
The correlation between \vcb and $\mu_\pi^2$ is also shown in Fig.~\ref{fig:cc}. 
The uncertainty on \vcb is dominated by the inclusive branching ratio. In the fit to the moments with the Configuration B, for the determination of \vcb we assume that the uncertainty on ${\cal B}(\Bsb \to X\ellm \bar{\nu}_\ell)$ will be reduced from $8\%$ to $3\%$.

\section{Outlook \& Conclusion}
\label{sec:conc}
We presented a proof-of-concept for a future precision determination of inclusive \Bs decays through a sum-over-exclusives technique. Doing so requires significantly more information on the \Bs decay modes than presently available. 

Specifically, it is important to improve the knowledge of semi-leptonic \Bs decays into \Dsstarstar states. This requires more precise determinations of the \Dsstarstar branching fractions into their non-leptonic and radiative final states, which are presently poorly known. The best prospects for improving the branching fractions are future measurements at BESIII~\cite{BESIII:2020nme} or at Belle II~\cite{Belle-II:2018jsg}. At hadron colliders, absolute branching fraction measurements are difficult to perform, as usually insufficient information about the initial state is available. One possibility was explored in \cite{Stone:2019nrd}, a similar approach could potentially be exploited for the \Dszero and \Dsonep states, \eg using \decay{\Bd}{\Dstarm\Dsp} decays, because the lightest excited charm states are known to decay exclusively via a \Dsp resonance. However, as of today, no measurement exists following this method.

In addition, a better understanding of the rate and shape of the non-resonant \decay{\Bsb}{D^{(*)}K\ellm \bar{\nu}_\ell} decays is required. From the experimental point of view, a refinement of the approach used by \lhcb in \cite{LHCb:2019fns} would allow a precise determination of the rate. This would also require improving the parametrization of the non-resonant component.   
The GR model used in our studies describes reasonably well the region close to the mass threshold. However, at higher masses, which is the region that mostly affects the higher-order hadronic moments, it remains to be seen if this parametrization is sufficiently precise. The (model-independent) framework recently developed in \cite{Gustafson:2023lrz} for $\bar{B}\to \Xc\ellm \bar{\nu}_\ell$ decays could therefore be useful when adapted to the \Bs case.

The approach described here could be extended to study hadronic moments of inclusive semileptonic decays of \Lb hadrons and other heavy hadrons. For these, the same OPE could be applied and similar expressions hold for the theoretical predictions (see \eg \cite{Colangelo:2020vhu} for a recent study of \Lb hadrons).  

The above strategy to improve the knowledge of the \Bs system will allow for a future precision measurement of inclusive \Bs decays. This enables an extraction of the HQE parameters of the \Bs meson for the first time from data. Comparing these to the parameters extracted from global fits of the inclusive $\bar{B} \to \Xc \ellm \bar\nu_\ell$ spectra serves as an important test of $SU(3)_F$ symmetry. Combined with a measurement of the total semileptonic rate this would result in a new determination of \vcb thereby shedding light on the inclusive/exclusive \vcb puzzle.

\section*{ Acknowledgments } 
We thank Matteo Fael, Paolo Gambino, Florian Herren, Thomas Mannel and Raynette van Tonder for useful discussions. 
The work of M.R. is supported in part by the Italian Ministry of Research (MIUR) under the grant PRIN 2022N4W8WR. The work of KKV is supported in part by the Dutch Research Council (NWO) as part of the project Solving Beautiful Puzzles (VI.Vidi.223.083) of the research programme Vidi.

\newpage
\bibliographystyle{LHCb.bst}
\bibliography{thebib}

\end{document}